\documentclass[a4paper,11pt]{article}
\pdfoutput=1 

\usepackage{jinstpub} 

\title{\boldmath Characterization of a novel pixelated Silicon Drift Detector (PixDD) for high-throughput X-ray astrophysics}

\author[a,b,1]{Y.~Evangelista,\note{Corresponding author.}}
\author[a]{F.~Ambrosino,}
\author[a,b]{M.~Feroci,}
\author[c]{P.~Bellutti,}
\author[d,e]{G.~Bertuccio,}
\author[c]{G.~Borghi,}
\author[f,g]{R.~Campana,}
\author[h]{M.~Caselle,}
\author[i,j]{D.~Cirrincione,}
\author[c]{F.~Ficorella,}
\author[k]{M.~Fiorini,}
\author[f,g]{F.~Fuschino,}
\author[d,e]{M.~Gandola,}
\author[l]{M.~Grassi,}
\author[f,g]{C.~Labanti,}
\author[l]{P.~Malcovati,}
\author[d,e]{F.~Mele,}
\author[a]{A.~Morbidini,}
\author[c]{A.~Picciotto,}
\author[j]{A.~Rachevski,}
\author[m]{I.~Rashevskaya,}
\author[d,e]{M.~Sammartini,}
\author[j]{G.~Zampa,}
\author[j]{N.~Zampa,}
\author[c]{N.~Zorzi}
\author[i,j]{and A.~Vacchi}

\affiliation[a]{INAF-IAPS,Via del Fosso del Cavaliere 100, I-00133 Rome, Italy}
\affiliation[b]{INFN sez. Roma 2, Via della Ricerca Scientifica 1, I-00133 Rome, Italy}
\affiliation[c]{Fondazione Bruno Kessler -- FBK, Via Sommarive 18, I-38123 Trento, Italy}
\affiliation[d]{Politecnico di Milano, Department of Electronics, Information and Bioengineering, Via Anzani 42, I-22100 Como, Italy}
\affiliation[e]{INFN sez. Milano, Via Celoria 16, I-20133 Milano, Italy}
\affiliation[f]{INAF-OAS Bologna, Via Gobetti 101, I-40129 Bologna, Italy}
\affiliation[g]{INFN sez. Bologna, Viale Berti-Pichat 6/2, I-40127 Bologna, Italy}
\affiliation[h]{Karlsruhe Institute of Technology, Hermann-von-Helmholtz-Platz 1, D-76344 Eggenstein-Leopoldshafen, Germany}
\affiliation[i]{University of Udine, Via delle Scienze 206, I-33100 Udine, Italy}
\affiliation[j]{INFN sez. Trieste, Padriciano 99, I-34127 Trieste, Italy}
\affiliation[k]{INAF-IASF Milano, Via Bassini 15, I-20100 Milano, Italy}
\affiliation[l]{University of Pavia, Department of Electrical, Computer, and Biomedical Engineering, and INFN Sez. Pavia, Via Ferrata 3, I-27100 Pavia, Italy}
\affiliation[m]{TIFPA-INFN, Via Sommarive 14, I-38123 Trento, Italy}

\emailAdd{yuri.evangelista@inaf.it}

\abstract{
Multi-pixel fast silicon detectors represent the enabling technology for the next generation of space-borne experiments devoted to high-resolution spectral-timing studies of low-flux compact cosmic sources. 
Several imaging detectors based on frame-integration have been developed as focal plane devices for X-ray space-borne missions but, when coupled to large-area concentrator X-ray optics, these detectors are affected by strong pile-up and dead-time effects, thus limiting the time and energy resolution as well as the overall system sensitivity. 
The current technological gap in the capability to realize pixelated silicon detectors for soft X-rays with fast, photon-by-photon response and nearly Fano-limited energy resolution therefore translates into the unavailability of sparse read-out sensors suitable for high throughput X-ray astronomy applications. 
In the framework of the ReDSoX Italian collaboration, we developed a new, sparse read-out, pixelated silicon drift detector which operates in the energy range 0.5--15 keV with nearly Fano-limited energy resolution ($\leq$150 eV FWHM @ 6 keV) at room temperature or with moderate cooling ($\sim$0~$^{\circ}$C to +20~$^{\circ}$C). In this paper, we present the design and the laboratory characterization of the first 16-pixel (4$\times$4) drift detector prototype (PixDD), read-out by individual ultra low-noise charge sensitive preamplifiers (SIRIO) and we discuss the future PixDD prototype developments.
}


\arxivnumber{XXXX.YYYYY} 

\collaboration[c]{on behalf of the ReDSoX collaboration}

\begin{document}
\maketitle

\flushbottom


\section{Introduction}
\label{sec:intro}

The study of high energy radiation from celestial objects is one of the most powerful diagnostic tools to access and understand the mechanisms underlying the most energetic and violent phenomena in the Universe. The X-ray range (0.5--10 keV) is particularly suitable for this investigation thanks to a mature detector and optics technology. Spectral and timing radiation signatures at these energies offer a direct access to plasma in environments hosting extreme conditions of gravity, density or magnetic field.
The need for high sensitivity spectroscopic and timing experiments has inspired in the last years intensive R\&D programs focused on the development of innovative, fast, pixelated detectors. 
As a matter of fact, when imaging X-rays, the size of the optics point spread function (PSF) is typically in the range of one millimetre or less for the usual focal lengths. In cases where imaging is not the primary scientific goal, the ideal pixel size is as small as required for an oversampling of the PSF, but as large as possible to reduce charge sharing and power consumption by reducing the number of required readout channels. In this energy range, silicon is probably the most suitable detector material, due to its quantum efficiency and advanced manufacturing technology.

\begin{figure}[!ht]
\centering
\includegraphics[height=0.35\textwidth]{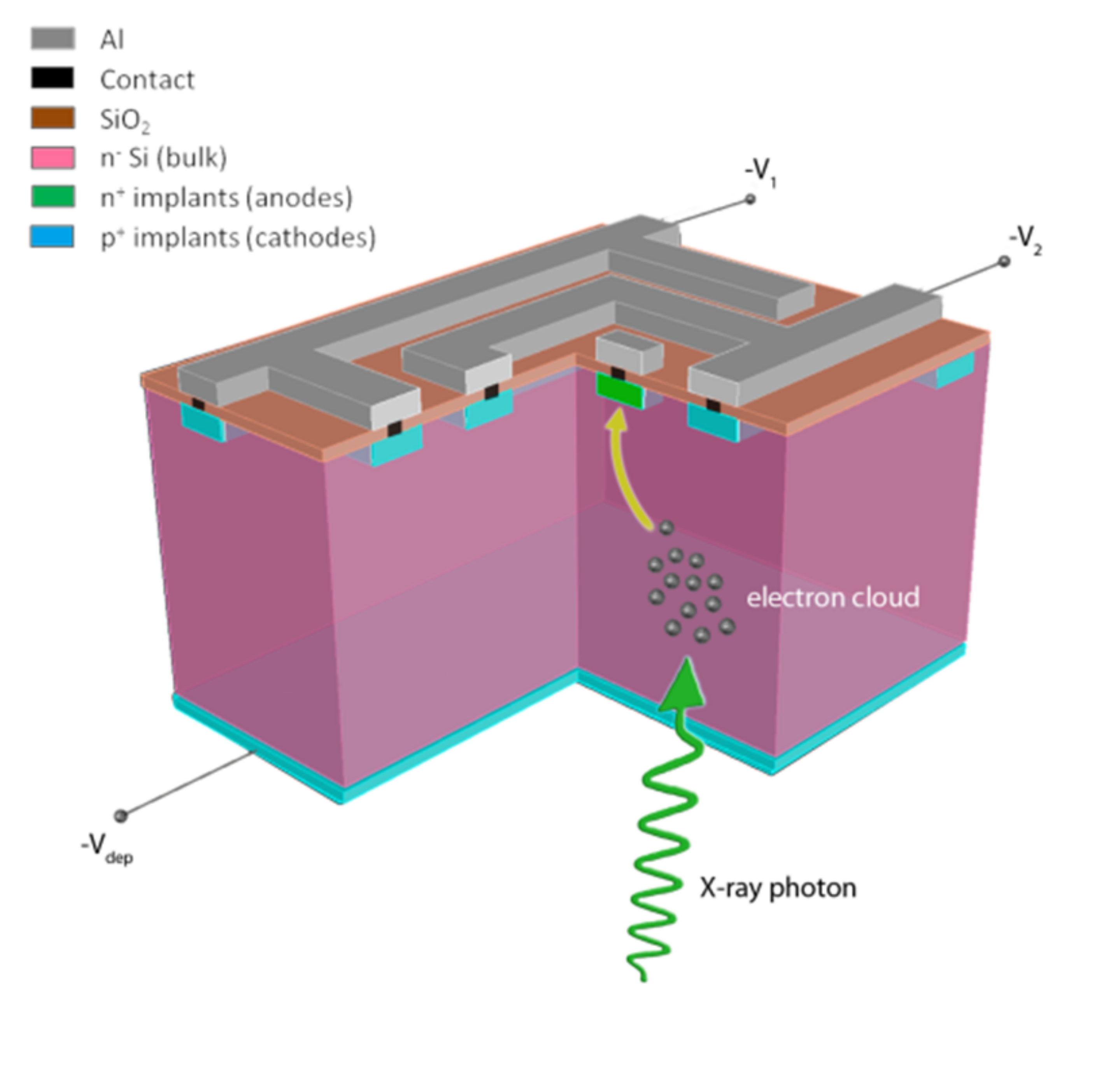}
\includegraphics[height=0.30\textwidth]{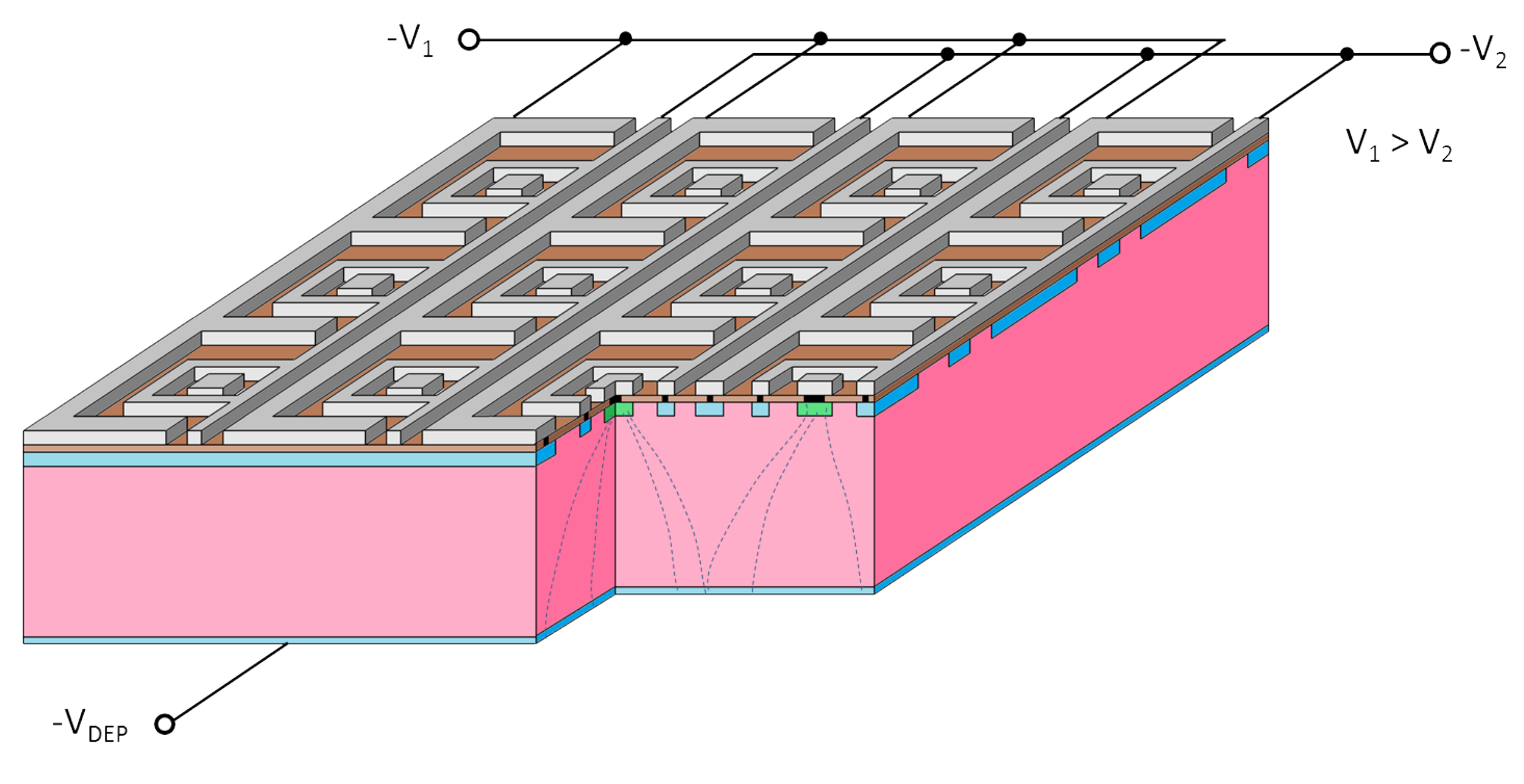}
\caption{Simplified single pixel layout (left) and multi-pixel layout (right) of the PixDD detector. See text for details.}
\label{fig:pixddlayout}       
\end{figure}

Different large pixel detectors have been proposed as focal plane sensor for X-ray spectral-timing study of astrophysical sources.
The widely used Si-PIN detectors show very interesting capabilities in terms of X-ray detection efficiency and sensor effective area. Their performance is however limited by the large anode capacitance (of the order of $10^3$ fF), which makes this kind of sensor not suitable for applications where a nearly Fano-limited energy resolution ($\leq$150 eV FWHM @ 5.9 keV) is required, especially if the experiment is characterized by a high counting rate (i.e. a short peaking time is required) with limited resources for detector cooling (i.e. large parallel noise) and read-out (small pixel size). 
Similarly, charge coupled devices (CCDs) with large pixel area suffer the same problems, even worsened by the integration of the leakage current between the read-out of two consecutive image frames (usually worked around with deep cooling, typically to temperatures lower than $-70$ $^{\circ}$C). Moreover, since the CCD is an integration-based imaging detector, it is not suitable for photon-by-photon spectral and timing information at high rates (for bright sources, due to pile-up): this is actually a requirement for the astrophysical applications described above.
Recently, Depleted p-channel Field Effect Transistors (DePFET \citep{Kemmer1987, Kemmer1990, Lutz2001}) have been developed and several prototypes have been manufactured. DePFET are excellent sensors for both space-borne imaging applications and high resolution X-ray spectroscopy when a frame rate around 10~kHz 
is required, as implemented for example in the Athena WFI high count rate capable sensor \citep{Meidinger2017}. An approach to a higher frame rate has been studied for the DePFET Sensor with Signal Compression camera (DSSC) at the European XFEL in Hamburg \citep{Porro2012}. Although in these devices all pixels are read out simultaneously at a rate of a few MHz, the high speed read-out requires a high power consumption and heat dissipation, making this scheme less than ideal for space-borne applications.

In this paper we present the design, production and laboratory characterization of a state-of-the-art pixelated detector prototype --- PixDD --- based on planar silicon technology. 
PixDD is aimed at the measurement of X-rays between 0.5 and 15 keV with a nearly "Fano-limited" spectral resolution, a timing resolution of few microseconds and a photon-by-photon fast read-out. 
The detector system exploits the superior noise characteristics typical of the Silicon Drift Detectors (SDD) and the ultra-low noise performance of a dedicated front-end electronics. 
The PixDD development builds on the state-of-the-art results achieved in Italy on both SDDs, with the combined design and manufacturing technology of INFN-Trieste and Fondazione Bruno Kessler (FBK, Trento), and the read-out electronics, with the unprecedented noise performance of the SIRIO charge preamplifier \citep{Bertuccio2007,Bertuccio2014} developed at Polytechnic of Milan.
This high-performance, pixelated silicon detector ($<$1 mm$^2$/pixel) has been designed aiming to imaging, timing and spectroscopic studies of astrophysical X-ray sources and will enable the development of a large-area detector with immediate application to spaceborne high-energy astrophysics experiments. 

\begin{figure}[!t]
\begin{center}
	\includegraphics[width=0.80\textwidth]{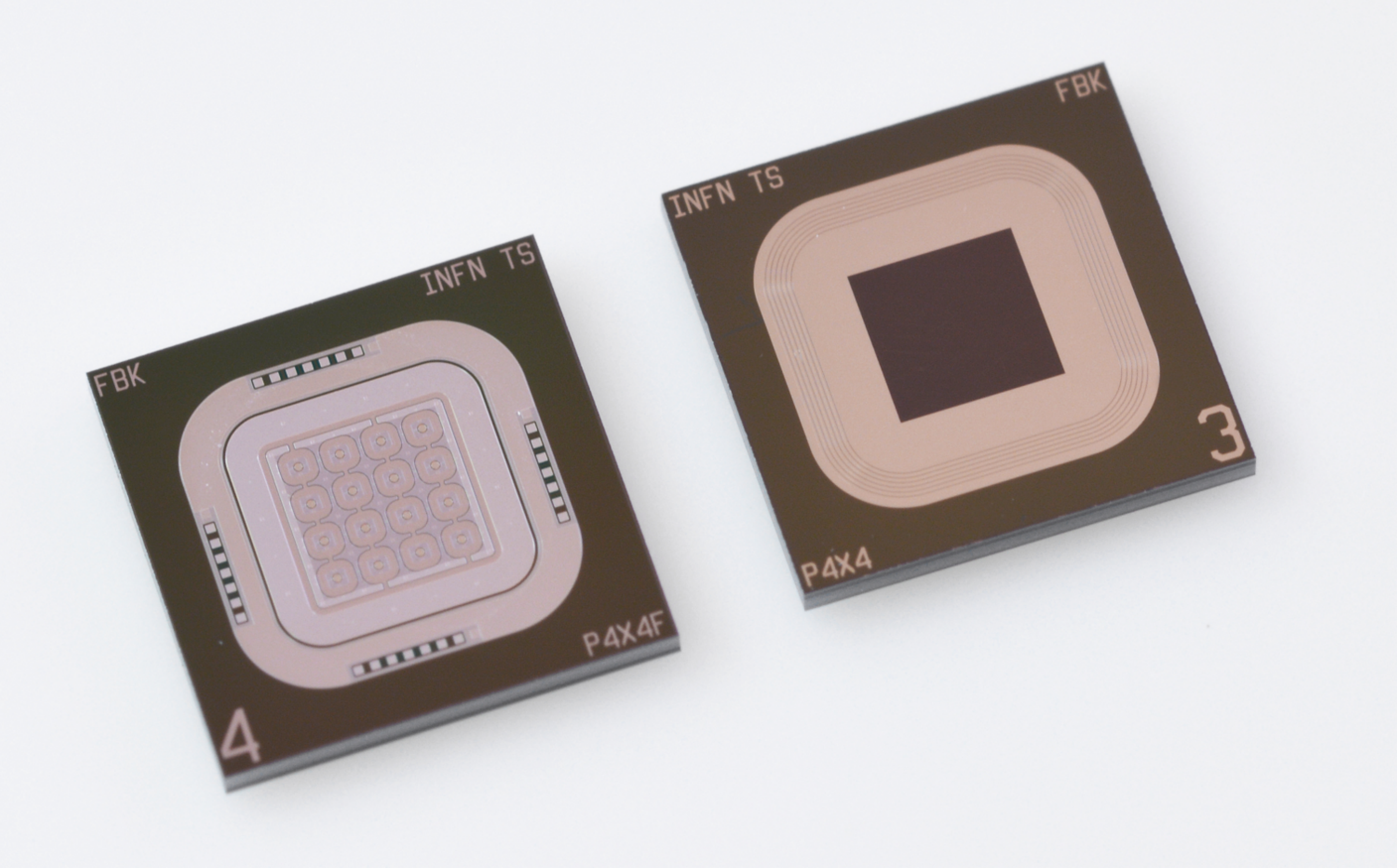}
	\caption{The first PixDD detector prototype. Left: n-side (anode side), Right: p-side (entrance window side).}	
	\label{fig:pixddphoto}       
\end{center}
\end{figure}

\begin{figure}[!t]
\begin{center}
	\includegraphics[width=0.80\textwidth]{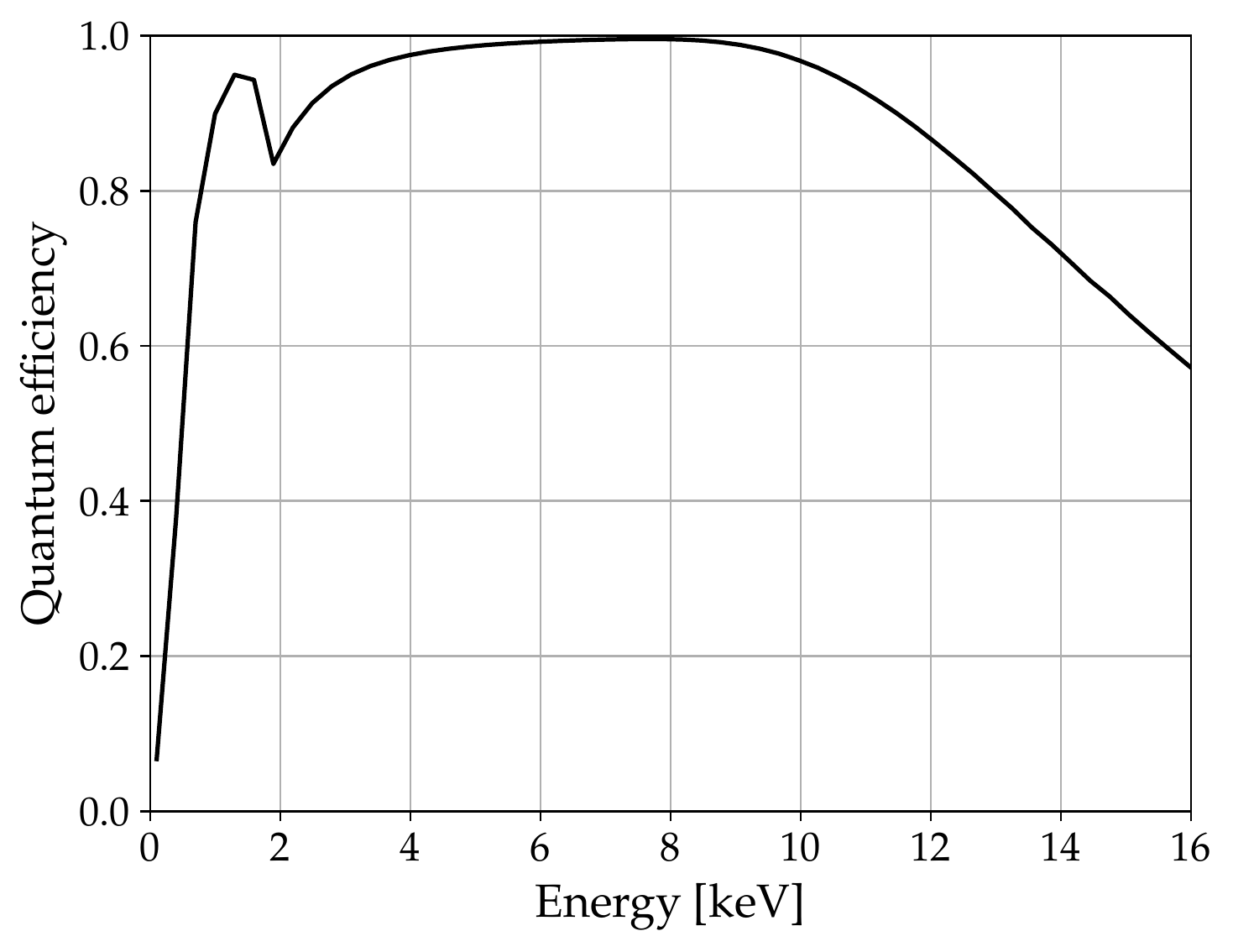}
	\caption{Estimated quantum efficiency of the PixDD detector for back-illumination.}	
	\label{fig:pixddqe}       
\end{center}
\end{figure}
   

\section{The Pixel Drift Detector -- PixDD}
\label{sec:pixdd}

The PixDD detector layout is aimed at reducing the total capacitance of the anode when compared with a classical pixel structure of the same size, thus minimizing the series noise contribution and enhancing the performance. The detector is composed of a matrix of small drift structures having a common shallow p$^+$ entrance window. On the anode side, two cathodes surrounding the collecting electrode provide a transversal field, which is designed to avoid regions in which the potential is slowly varying and the drift velocity is therefore very small.
In Figure~\ref{fig:pixddlayout} the simplified PixDD detector layout is shown. The two drift cathodes (in grey) are biased from the external circuitry without the need for integrated resistors. This allows to minimize the detector power consumption with respect to the use of an integrated voltage divider for each detector cell.
The shallow p$^+$ entrance window (on the bottom in Figure~\ref{fig:pixddlayout}) is biased at a negative potential with respect to the anodes ($\mathrm{-85\,V \leq V_{dep} \leq -75\,V}$) to ensure the full depletion of the detector bulk.

The first PixDD detector prototype (Figure~\ref{fig:pixddphoto}) has been fabricated  by FBK on a 450~$\mathrm{\mu m}$ thick n-type, floating-zone (FZ) silicon substrate with a resistivity of $\mathrm{9 \, k\Omega \cdot cm}$. The detector is made of a total of 16 square pixels $\mathrm{500~\mu m \times 500~\mu m }$ wide, arranged in 4 rows by 4 columns. By design, the anode capacitance is about 30~fF.

The shallow p$^+$ implant of the entrance window guarantees a good detector quantum efficiency for low energy photons. As shown in Figure~\ref{fig:pixddqe}, the calculated efficiency is larger than 50\% for photons of energy $E\geq500$~eV, reaching values well above 80\% in the 1~keV--8~keV energy band.

\section{PixDD read-out system}
\label{sec:readout}

In order to characterize the novel PixDD detector an ultra low-noise charge preamplifier has been selected to ensure the minimum contribution of the front-end electronics to the detector noise and energy resolution. All the tests described in this paper have been performed with the SIRIO 3 charge sensitive preamplifier (CSA) developed at the Polytechnic of Milan \citep{Bertuccio2014} to achieve a nearly Fano limited resolution performance. The SIRIO performance (ENC of 1.3 e$^{-}$ r.m.s. at room temperature and 0.89 e$^{-}$ r.m.s. at $-30$~$^{\circ}$C) have been also demonstrated  with a 13 mm$^2$ hexagonal SDD and a  monolithic multi-element SDD \citep{Bertuccio2015, Bertuccio2016, Bufon2018}, both designed and produced by INFN-Trieste and FBK in the framework of the ReDSoX collaboration\footnote{\url{http://redsox.iasfbo.inaf.it}}.

\begin{figure}[!t]
\begin{center}
	\includegraphics[width=0.70\textwidth]{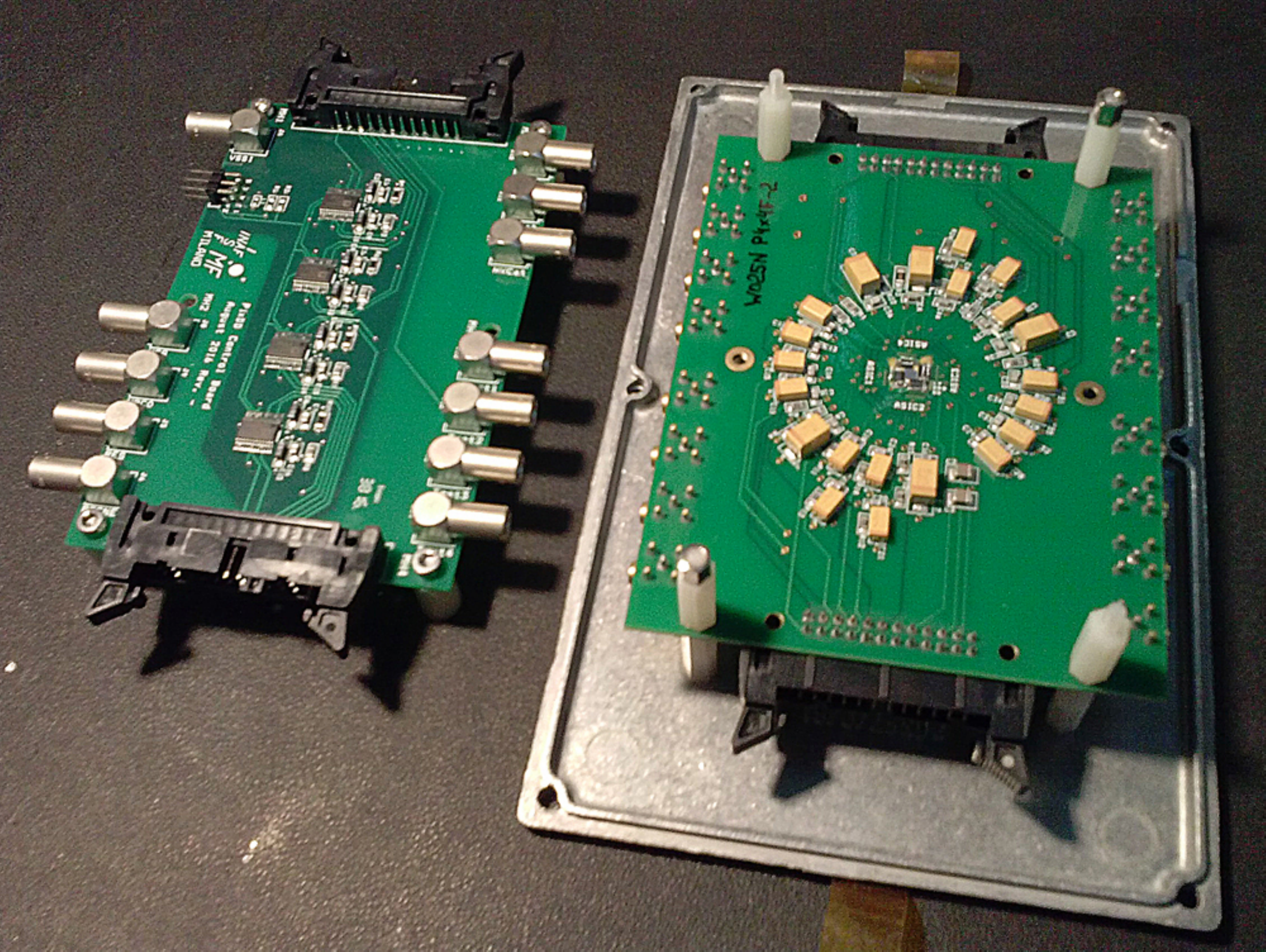}
	\caption{PixDD control board (PixDD-CB, left) and PixDD detector board (PixDD-DB, right) developed at the Istituto di Astrofisica Spaziale e Fisica Cosmica (IASF) in Milan.}	
	\label{fig:pixddpcb}       
\end{center}
\end{figure}

To simplify the interconnection scheme between the CSA and the 16 channels of the PixDD detector, a custom 4-channel SIRIO preamplifier (SIRIO 3.4) has been designed and built in 0.35~$\mu$m AMS CMOS technology. The four channels share the CSA biases and the reset signals, while the inputs, the test injection pads and the feedback network fine-tuning voltages (V$_{\mathrm{ssi}}$) are individual for each channel. The overall dimensions of the 4-channel preamplifier chip are $\mathrm{2300~\mu m \times 1820~\mu m}$.

Two custom PCBs, namely PixDD control board (PixDD-CB) and PixDD detector board (PixDD-DB), have been designed and manufactured by IASF Milano and IAPS Rome to integrate the PixDD detector with the SIRIO 3.4 chips. A picture of the two PCBs is shown in Figure~\ref{fig:pixddpcb}.
The PixDD-CB provides bias line distribution for both the detector and the CSA chips and allows for the fine regulation of the 16 feedback network fine-tuning voltages. The setting of the V$_{\mathrm{ssi}}$ voltages is performed by means of digital potentiometers (AD5263 by Analog Devices) controlled with $\mathrm{I^2C}$ interface.
The PixDD-DB provides effective filtering of the bias lines, impedance matching for the reset and test pulse signals, detector and read-out chip mounting and interconnection.
Moreover, an analogue temperature transducer (Analog Devices AD590KF) placed in close proximity of the detector die allows for a direct monitoring of the detector temperature. 
Input and output analogue signals are available through a set of $\mathrm{50\,\Omega}$ MCX connectors placed on the PixDD-DB perimeter, while the interconnection between the two PCBs is accomplished by means of ribbon flat cables.

\begin{figure}[!t]
\centering
\includegraphics[width=0.90\textwidth]{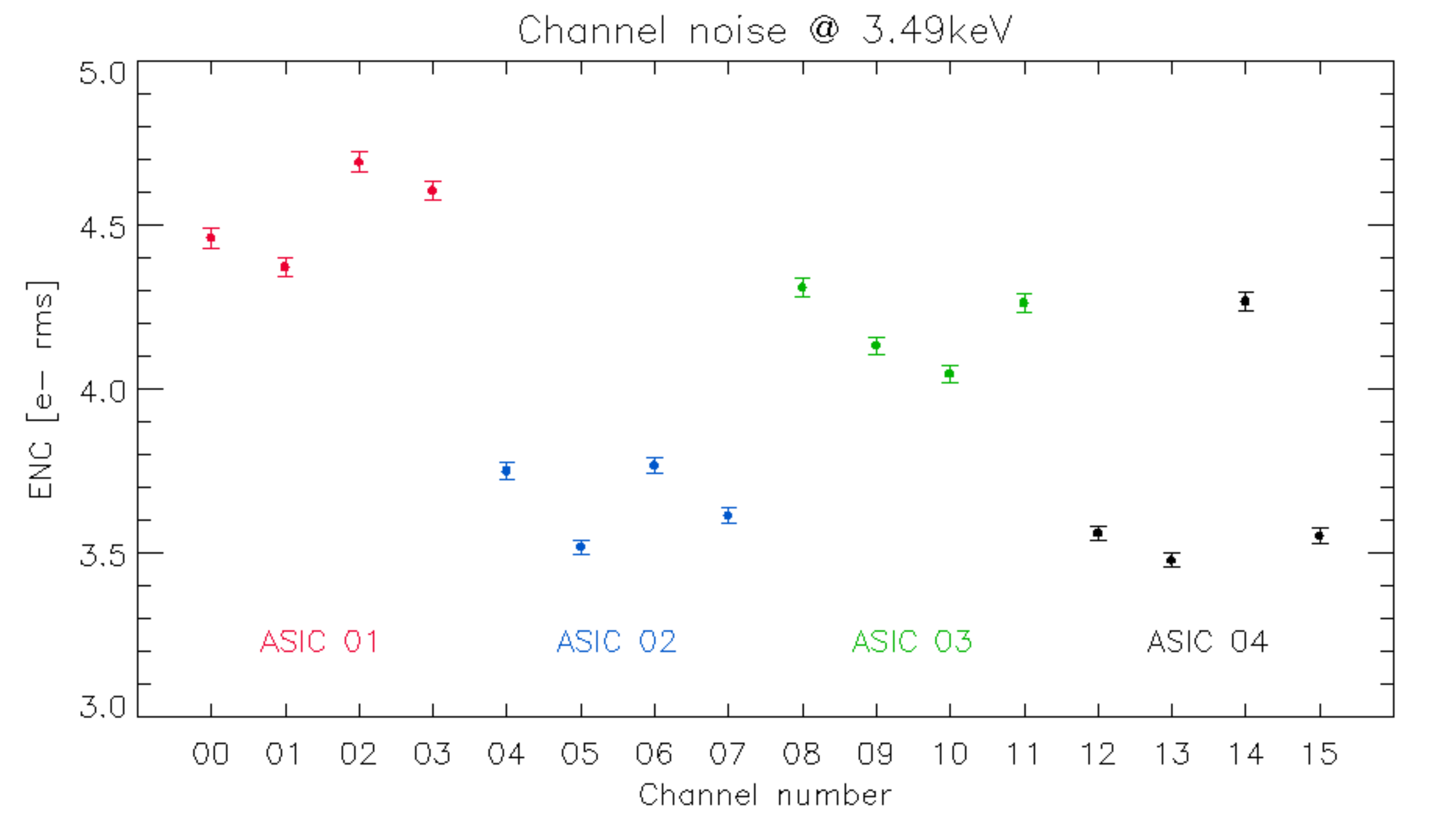}
\caption{Equivalent noise charge (ENC) measured with a pulser at 3490~eV (equivalent energy in silicon) for the 16 channels contained in the 4 preamplifier chips.}
\label{fig:sirioperf}       
\end{figure}


\subsection{Front-end electronics characterization}
\label{subsec:feetests}
The performance of the front-end system (FEE) has been tested and characterized in the INAF-IAPS laboratory after the integration of the SIRIO 3.4 ASICs with the PixDD-DB but before the interconnection of the PixDD anodes with the read-out input pads. 
The system biases were provided by standard linear power supplies, filtered by 2-stage RC networks and integrated EMI filters. The injection of charge pulses was made by means of voltage steps applied to the SIRIO 3.4 test capacitances and generated by an Arbitrary Function Generator (Rigol DG5251). A passive 40--60 dB attenuator was used to maximize the signal to noise ratio of the injected voltage steps. A second Arbitrary Function Generator (Tektronix AFG3022B) was used to generate the periodic digital signal for the preamplifier reset.
After a further amplification ($\mathrm{G_v = 10}$), carried out by means of a custom AC coupled amplifier, the SIRIO 3.4 output signals were processed and acquired with a Digital Pulse Processor (Amptek DP5).

In order to characterize each system channel in terms of noise and functionality, the 16 CSAs have been individually stimulated with different test-pulse amplitudes, ranging from 170~eV to 3490~eV (equivalent energy in silicon). An average equivalent noise charge (ENC) of $4.0 \pm 0.4$ e$^-$ r.m.s. has been measured at $\mathrm{+21\,^{\circ}C}$ in the 500~eV--3490~eV energy range (40 dB attenuation applied to the test signal). A significantly better ENC of $2.4 \pm 0.2$ e$^-$ r.m.s has been obtained between 170~eV and 500~eV (attenuator set to 60~dB), most likely indicating the presence of a non-negligible noise contribution from the Arbitrary Function Generator. Although the measured ENC at room temperature is still substantially larger than the value of 1.3~$\mathrm{e^{-}}$ r.m.s. measured with the single channel SIRIO \citep{Bertuccio2016}, the overall FEE performance has been considered very promising to fulfil the energy resolution requirement at room temperature ($\leq$150~eV FWHM at 5.9 keV) of the integrated detector system.

Figure~\ref{fig:sirioperf} shows the distribution of the ENC for the 16 channels contained in the 4 preamplifier chips. It is worth noticing that for the conversion of the test-pulse amplitude in electrons the nominal value of the integrated test capacitance has been used, thus neglecting the expected spread related to the production process (typically of the order of 20\%).

\begin{figure}[!p]
\centering
\includegraphics[height=45mm]{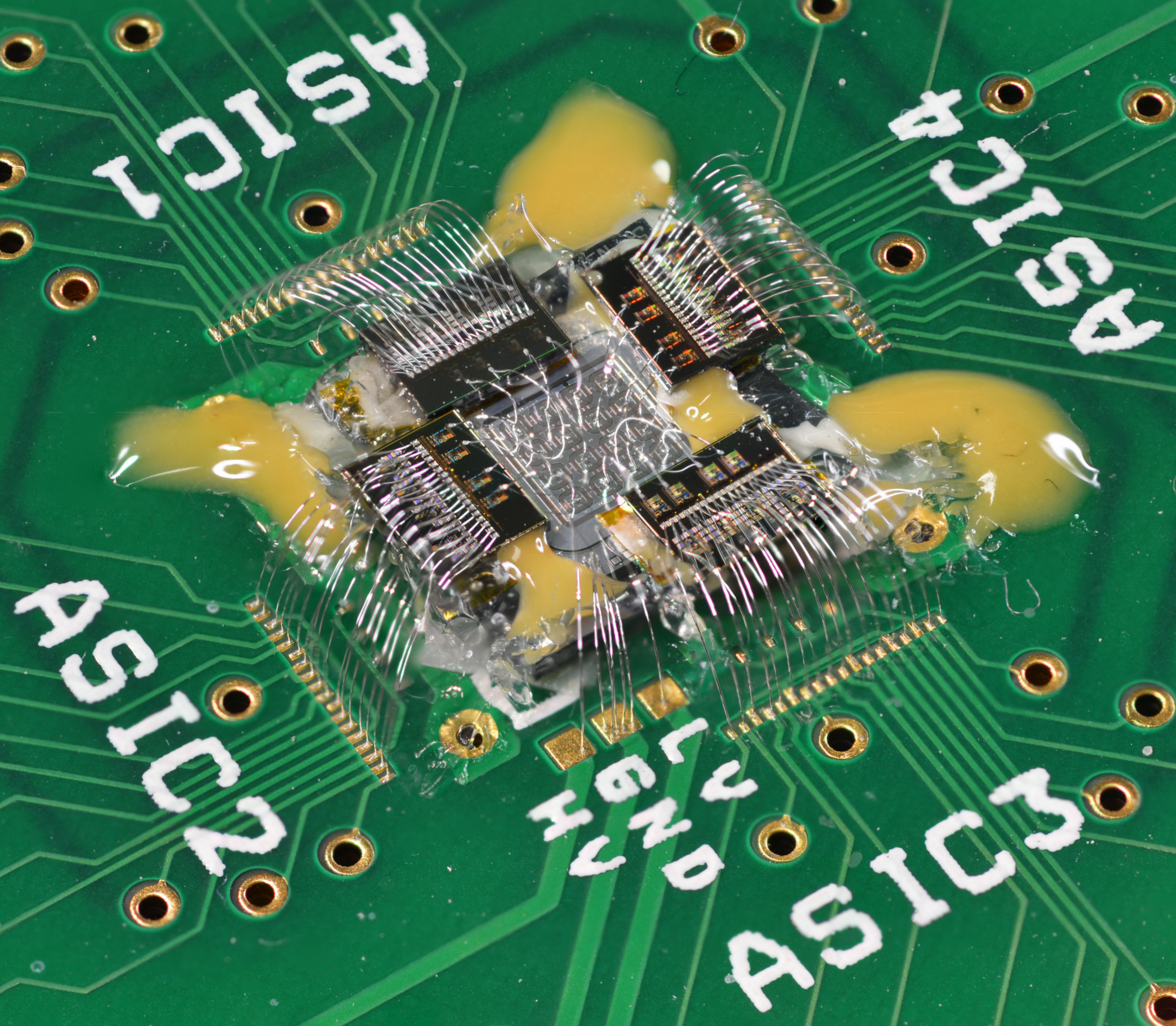}
\includegraphics[height=45mm]{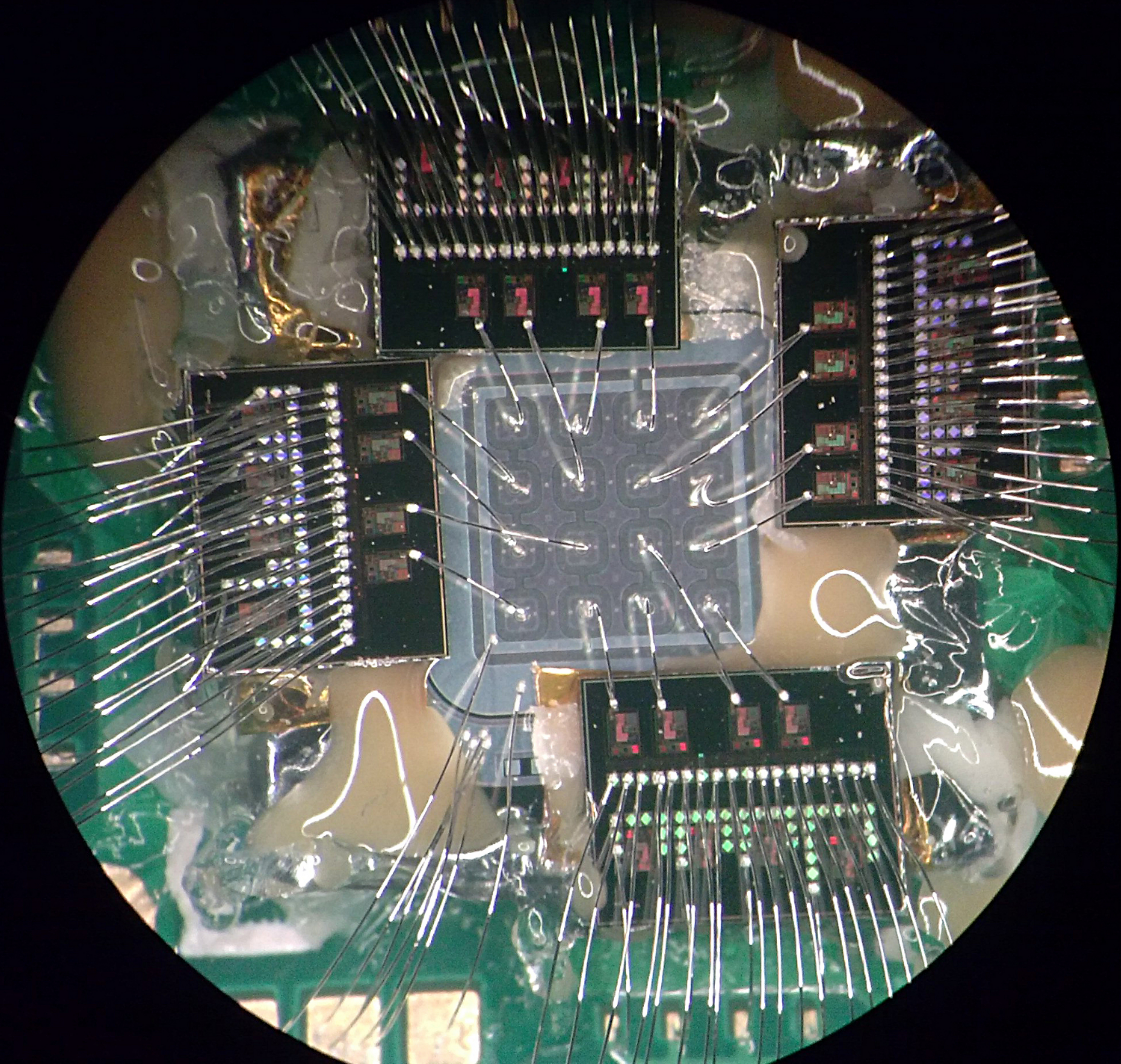}
\includegraphics[height=45mm]{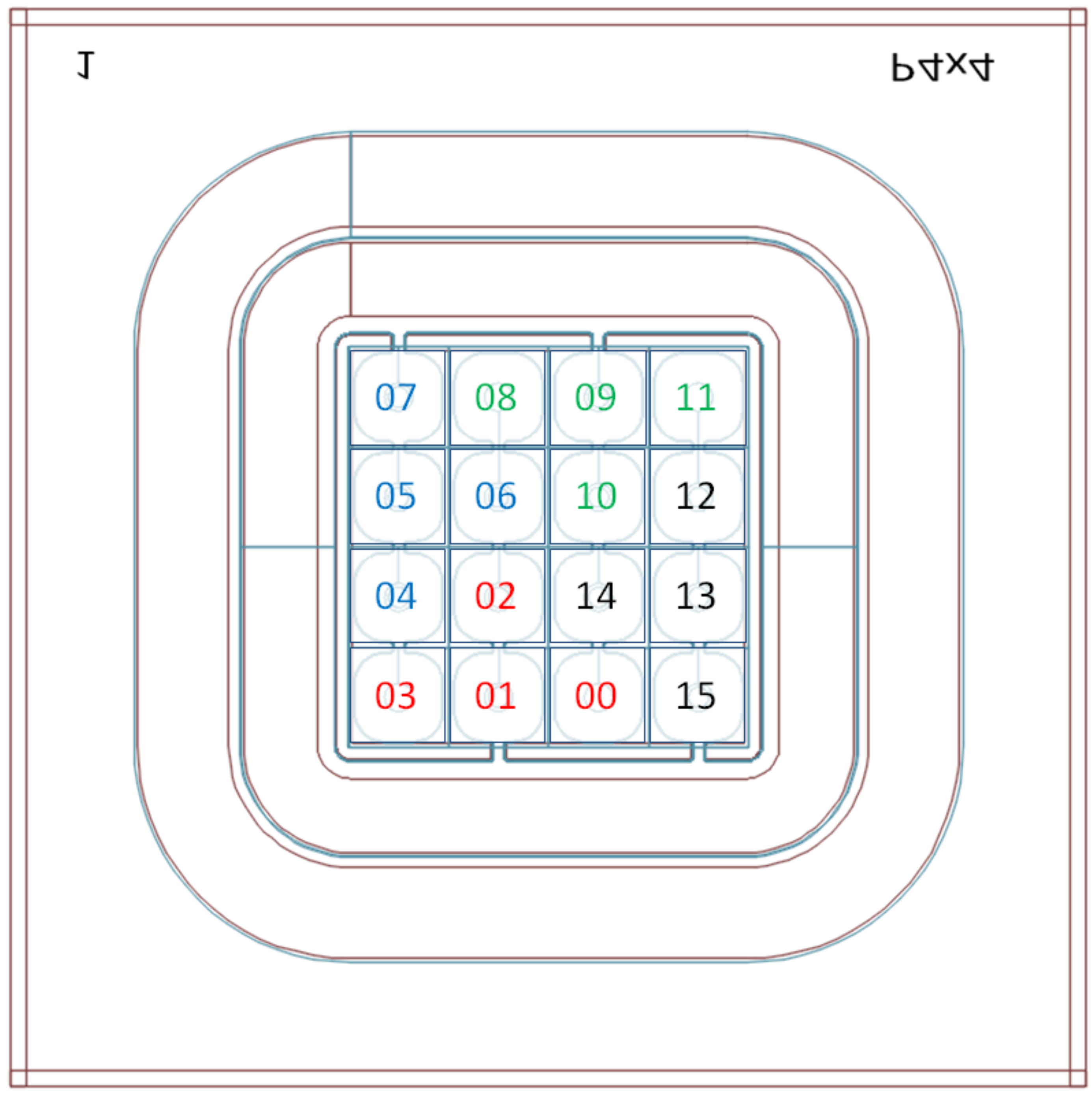}
\caption{Left and central panels: the PixDD detector wire bonded to the 16 SIRIO 3.4 channels. Right: correspondence of SIRIO 3.4 channels with PixDD pixels as seen from the p-side detector window. }
\label{fig:pixddbond}       
\end{figure}

\begin{figure}[!p]
\centering
\includegraphics[width=0.9\textwidth]{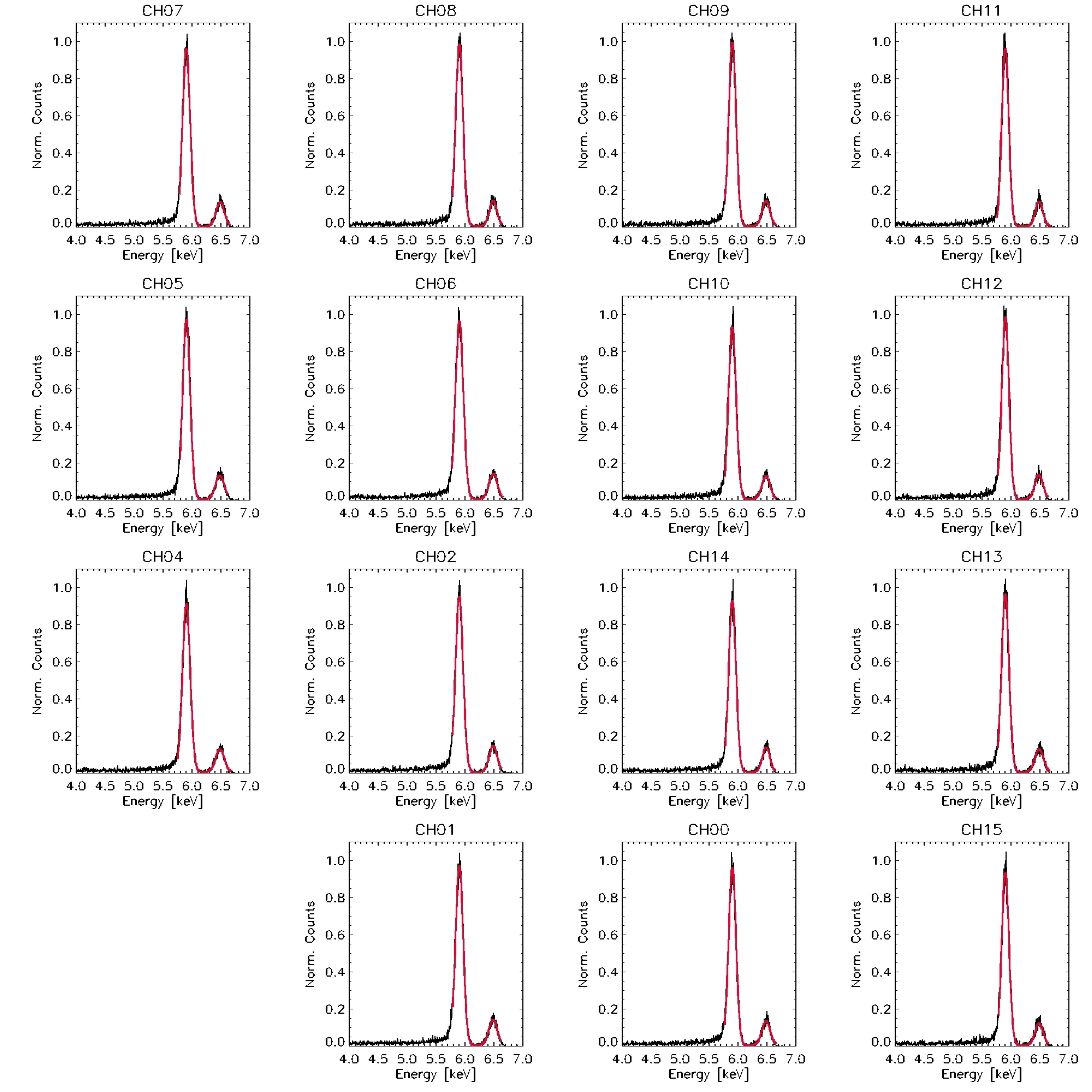}
\caption{$^{55}$Fe energy spectra acquired with 1~$\mu \mathrm{s}$ peaking time. CH03 (on the bottom-left corner of the detector) has not been acquired due to  problems during the CSA--anode interconnection.}
\label{fig:55fespectra}       
\end{figure}


\section{PixDD detector characterization}
\label{sec:characterization}

After the completion of FEE system electrical characterization described in Section~\ref{subsec:feetests}, the 16 PixDD anodes have been wire-bonded at the INFN Trieste laboratories to the relevant CSA input pads. The adopted bonding scheme, shown in Figure~\ref{fig:pixddbond}, was designed to minimize the bonding length and thus the stray capacitance contribution of the wire bondings to the system series noise.
Fifteen out of sixteen CSA channels were successfully wire-bonded to the PixDD anodes, while channel number 3 (CH03 hereafter) experienced an interconnection problem during the wire bonding procedure of the SIRIO 3.4 input pad, resulting unusable for all the following measurements. 
The complete rework of the PixDD-DB (i.e. removal of the damaged ASIC, glueing of a new SIRIO 3.4 ASIC, wire bonding) needed to recover CH03 functionality was not considered a high priority activity taking into account the prototypal nature of the detector. 

The physical correspondence between pixels and FEE channels after the interconnection is shown in the rightmost panel of Figure~\ref{fig:pixddbond}. Different number colours refer to different SIRIO 3.4 ASICs (red: ASIC 1, blue: ASIC 2, green: ASIC 3 and black: ASIC 4).
The integrated detector has been enclosed in a light-tight aluminium box to provide proper shielding. All the measurements presented in this Section have been performed 
at a detector temperature of $\mathrm{+27\pm 1\,^{\circ}C}$, with the box continuously exposed to a nitrogen flow in order to maintain a dry atmosphere.


\subsection{Functional tests}
\label{subsec:functional}

In the environmental conditions described above, a first functional test has been carried out by illuminating the PixDD detector with a $\mathrm{^{55}Fe}$ radioactive source. The data were acquired and processed individually on all the pixels by means of the same set-up used for the FEE characterization (Section~\ref{subsec:feetests}). The energy spectra, obtained with  a peaking time of 1~$\mathrm{\mu s}$ at $\mathrm{+27\,^{\circ}C}$, are displayed in Figure~\ref{fig:55fespectra}. The FWHM energy resolution of all the channels lies between 141~eV and 156~eV, which corresponds to an ENC of 8.9--11.8~$\mathrm{e^-}$ r.m.s.

The spectra in Figure~\ref{fig:55fespectra} have also been used to properly calibrate the gain (i.e. volts--electronvolts conversion) of the full electronic chain of the 15 interconnected channels. Using these values, the detector leakage current ($\mathrm{I_{leak}}$) has been evaluated pixel-by-pixel exploiting the measurement of the output ramp slope of the preamplifiers.
Table~\ref{tab:pixddileak} reports the $\mathrm{I_{leak}}$  measured at $\mathrm{+27\,^{\circ}C}$ and, for reference, the leakage current  values normalized to $\mathrm{+20\,^{\circ}C}$, estimated taking into account the exponential dependence of $\mathrm{I_{leak}}$ on temperature \citep{Spieler2005}.

\begin{table}[!h]
\caption{PixDD pixel leakage current ($\mathrm{I_{leak}}$) measured at $\mathrm{+27\,^{\circ}C}$ and rescaled at $\mathrm{+20\,^{\circ}C}$ }
\label{tab:pixddileak}
\centering
\begin{tabular}{ccc}
\hline\noalign{\smallskip}
PixDD & $\mathrm{I_{leak}}$ at $\mathrm{+27\,^{\circ}C}$ & $\mathrm{I_{leak}}$ at $\mathrm{+20\,^{\circ}C}$ \\
channel &  [pA] &  [pA]  \\
\noalign{\smallskip}\hline\noalign{\smallskip}
00	&	2.51	&	1.43	\\
01	&	2.40	&	1.36	\\
02	&	2.72	&	1.55	\\
03	&	N/A	    &	N/A	    \\
04	&	2.52	&	1.43	\\
05	&	2.50	&	1.42	\\
06	&	2.89	&	1.64	\\
07	&	2.62	&	1.49	\\
08	&	2.76	&	1.57	\\
09	&	2.67	&	1.52	\\
10	&	2.85	&	1.62	\\
11	&	2.43	&	1.38	\\
12	&	2.52	&	1.43	\\
13	&	2.54	&	1.44	\\
14	&	2.67	&	1.52	\\
15	&	2.34	&	1.33	\\
\noalign{\smallskip}\hline
\end{tabular}
\end{table}

The average $\mathrm{I_{leak}}$  measured on the 15 pixels at $\mathrm{+27\,^{\circ}C}$ is 
2.60~pA  (which corresponds to 1.48~pA at $\mathrm{+20\,^{\circ}C}$), 
with a standard deviation of less than 6.5\% on the whole detector, thus demonstrating the quality of the detector design and fabrication process. It is worth noticing that the current FBK production process has demonstrated even better results in terms of bulk leakage current, as reported for example in \citep{Bertuccio2016}.


\subsection{PixDD pixel characterization }
\label{sec:pixelcharacterization}

\begin{figure}[!h]
\centering
\includegraphics[width=0.8\textwidth]{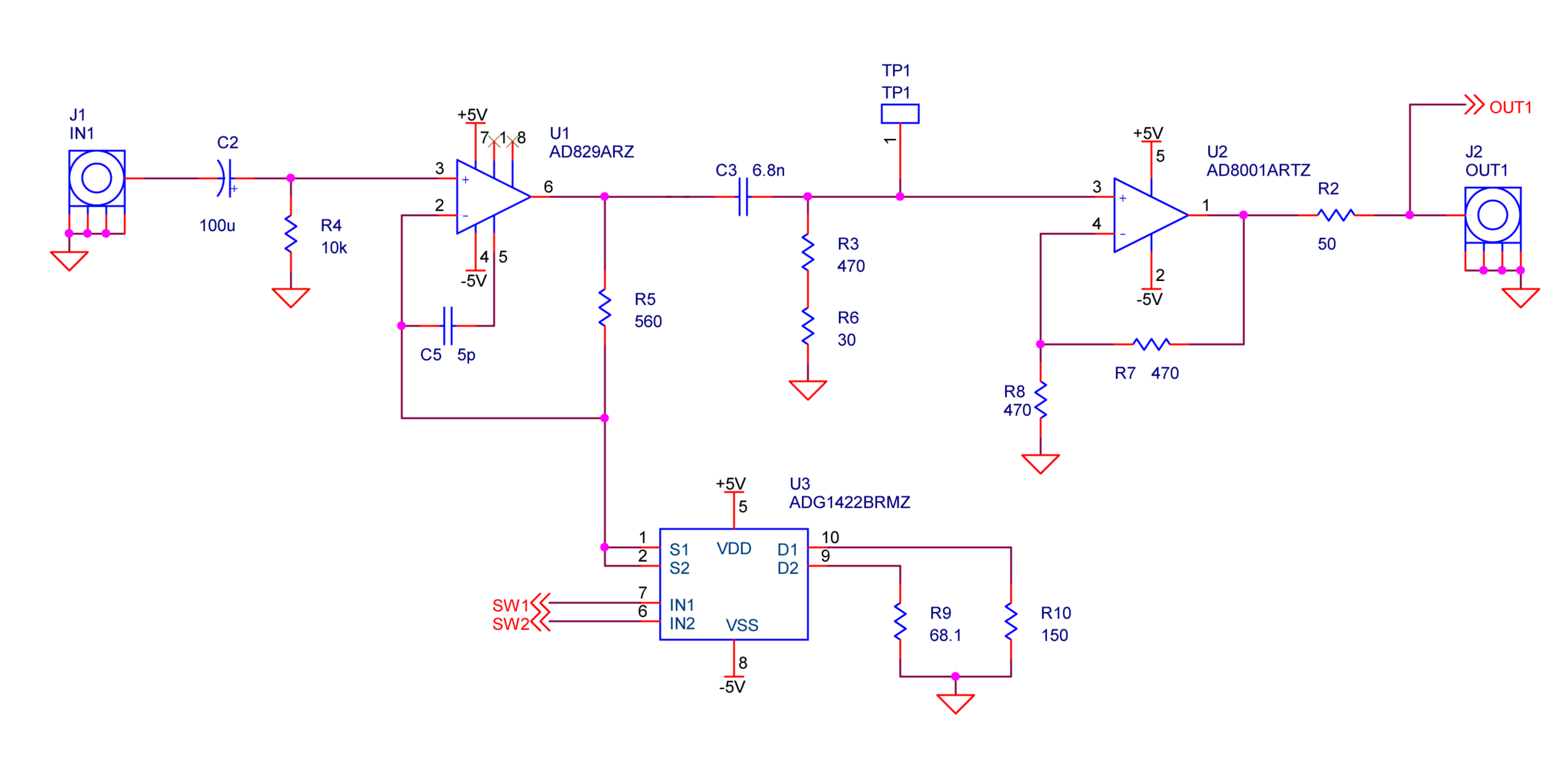}
\caption{Simplified schematic of the custom built, 16-channels amplifying (selectable gain of $\mathrm{A_v =}$ 4.7, 9.2, and 13.0) and pre-filtering (high-pass CR network with $\mathrm{3.4\,\mu s}$ time constant) board. Only one channel is shown.}
\label{fig:bufsch}       
\end{figure}

The signal conditioning and digitization system used so far for the PixDD characterization is limited to the acquisition of a single channel. To overcome this limitation, a multi-channel digitization system, based on the CAEN DT4750 fast digitizer, has been built. A custom 16-channels interface board has been designed and manufactured to provide proper amplification of the signals (with selectable gain  $\mathrm{A_v =}$ 4.7, 9.2, and 13.0) and  analogue pre-filtering by means of a high-pass CR network with $\mathrm{3.4\,\mu s}$ time constant. The simplified schematic of one pre-filter channel is shown in Figure~\ref{fig:bufsch}.
The signals have been digitized by the CAEN DT4750 digitizer at $\mathrm{62.5\,MS \cdot s^{-1}}$ with 12 bit resolution, controlled by a specifically developed LabView interface. The pulse-height analysis has been performed off-line by means of a digital trapezoidal filter (see, e.g., \citep{Guzik2013} and references therein). 
In presence of a trigger signal, provided by a digital threshold discriminator implemented in the DT4750 firmware, all the connected channels are sampled simultaneously and the waveforms are stored in a FITS binary table. This multi-channel acquisition system allowed for an in-depth characterization of the detector response. 


\subsubsection{Pencil beam characterization at 4.5~keV}
\label{sec:xray}

\begin{figure}[!t]
\centering
\includegraphics[width=0.9\textwidth]{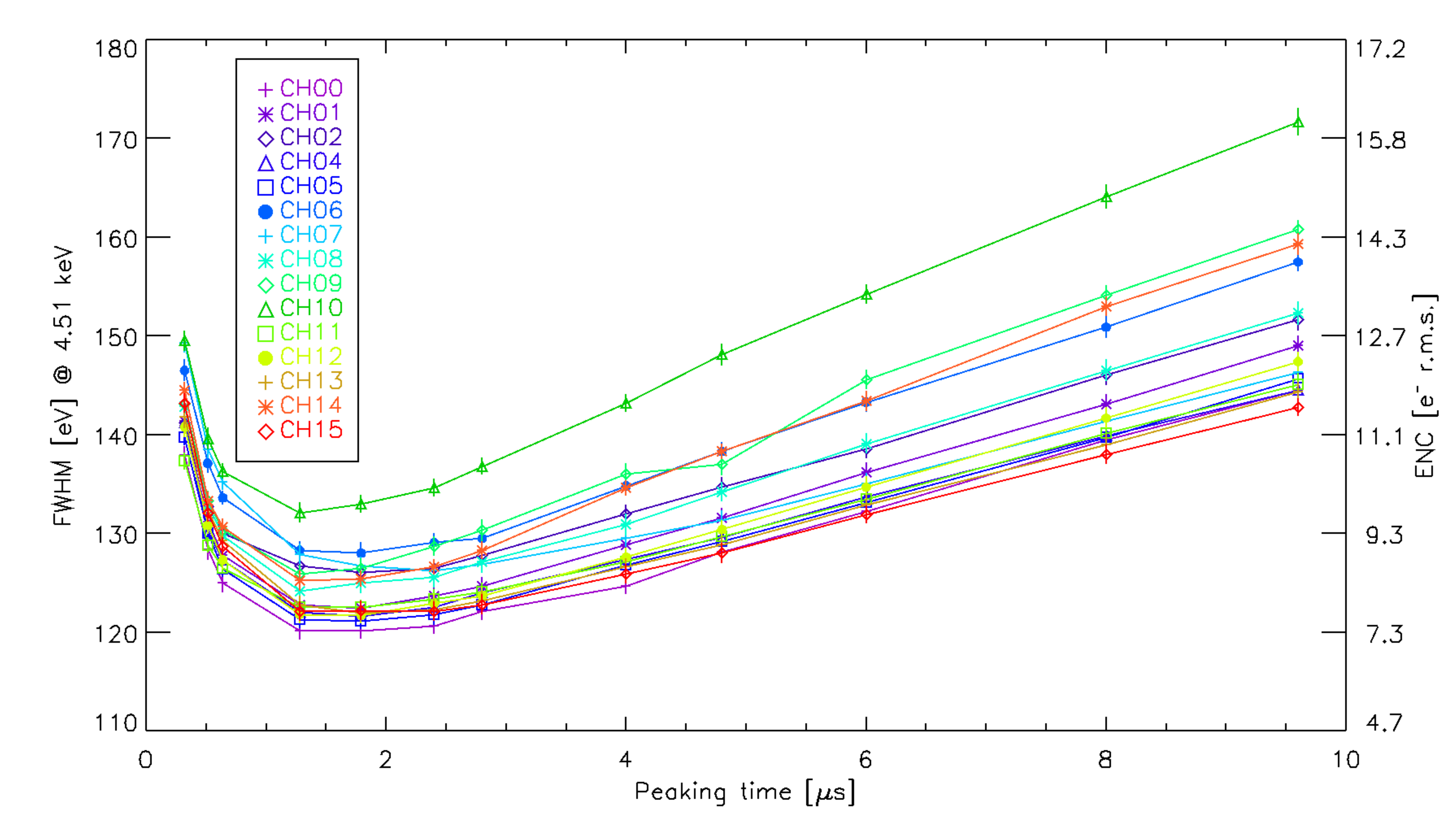}
\caption{Energy resolution at 4.51~keV and the corresponding equivalent noise charge (ENC) as a function of the peaking time for the 15 read-out channels.}
\label{fig:Tishaping}       
\end{figure}

\begin{figure}[!t]
\centering
\includegraphics[width=0.9\textwidth]{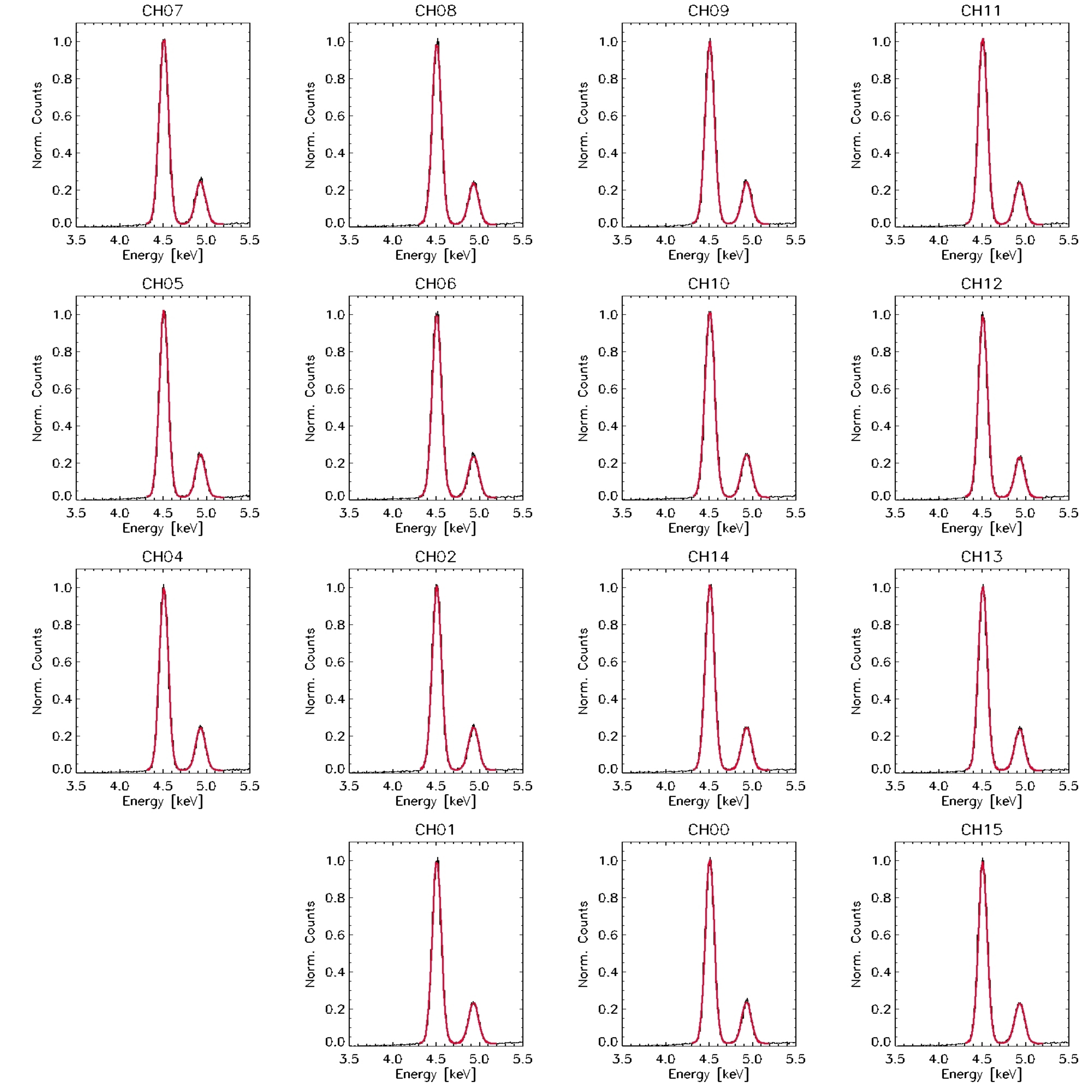}
\caption{Fitted energy spectra of a Ti-anode X-ray tube}
\label{fig:Tispectra}       
\end{figure}

A dedicated measurement campaign was carried out in the X-ray facility of the INAF-IAPS Rome laboratory \citep{Muleri2008} with the aim of characterizing the noise properties of the integrated PixDD-FEE system. The facility was equipped with a commercial Ti-anode X-ray tube and a custom collimation system to produce a $\mathrm{\sim 40\,\mu m}$ FWHM wide X-ray spot on the detector surface. A three-axis motorized linear stage system was used to position the X-ray spot in the centre of each PixDD pixel with micrometric precision.

Each PixDD pixel has been exposed to more than $7.3\times10^4$~counts in the 2--15~ keV energy band, with a count rate of about $\mathrm{175\, counts\cdot s^{-1}}$  (corresponding to a flux of  $\mathrm{7.0\times10^4\, counts\cdot\,cm^{-2}\cdot\,s^{-1}}$), for a total of $1.1\times10^6$~counts collected by all the 15 working channels.
For each channel, the acquired signal waveforms were digitally shaped with different peaking times ranging from $\mathrm{0.32\,\mu s}$ to $\mathrm{9.6\,\mu s}$ and the resulting energy spectra were fitted with two Gaussian functions centred at 4.51~keV and 4.93~keV (Titanium $\mathrm{K_{\alpha}}$ and $\mathrm{K_{\beta}}$ fluorescence lines) to estimate the energy resolution and thus the single-channel noise.

Figure~\ref{fig:Tishaping} shows the energy resolution and the equivalent noise charge  at 4.51~keV (Ti $\mathrm{K_{\alpha}}$) as a function of the peaking time for the 15 channels. All channels, except the slightly noisier CH10, show a FWHM energy resolution comprised between 120.1~eV and 128.0~eV at the optimal peaking time of $\mathrm{1.8\,\mu s}$, which corresponds to an ENC between $\mathrm{7.0\,e^-}$ and $\mathrm{8.7\,e^-}$ r.m.s.
The fitted energy spectra for all the channels are shown in Figure~\ref{fig:Tispectra}.


\subsubsection{X-ray characterization with $\mathrm{^{55}Fe}$ radioactive source}

\begin{figure}[!p]
\centering
\includegraphics[width=0.7\textwidth]{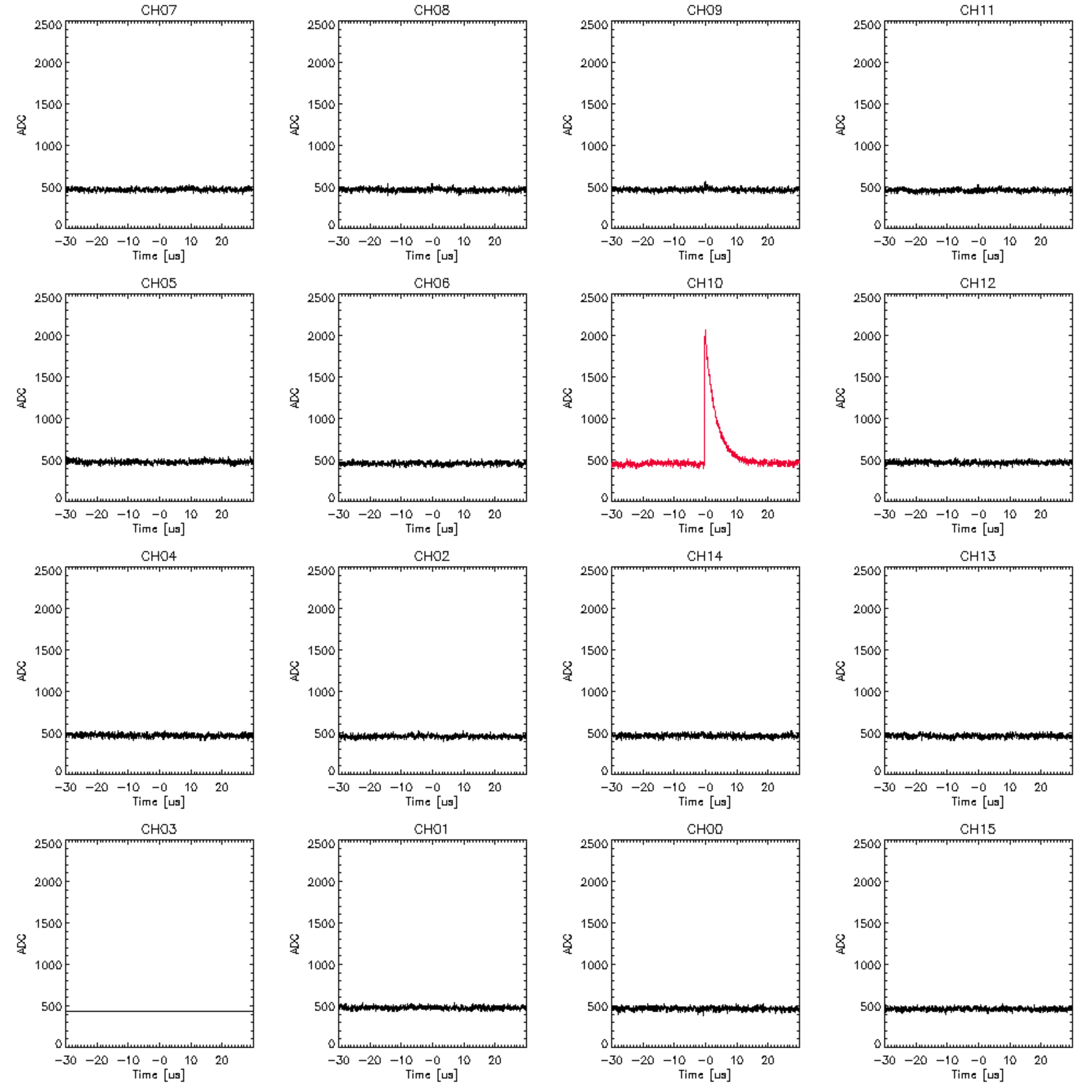}
\vspace{0.75cm}
\\
\includegraphics[width=0.7\textwidth]{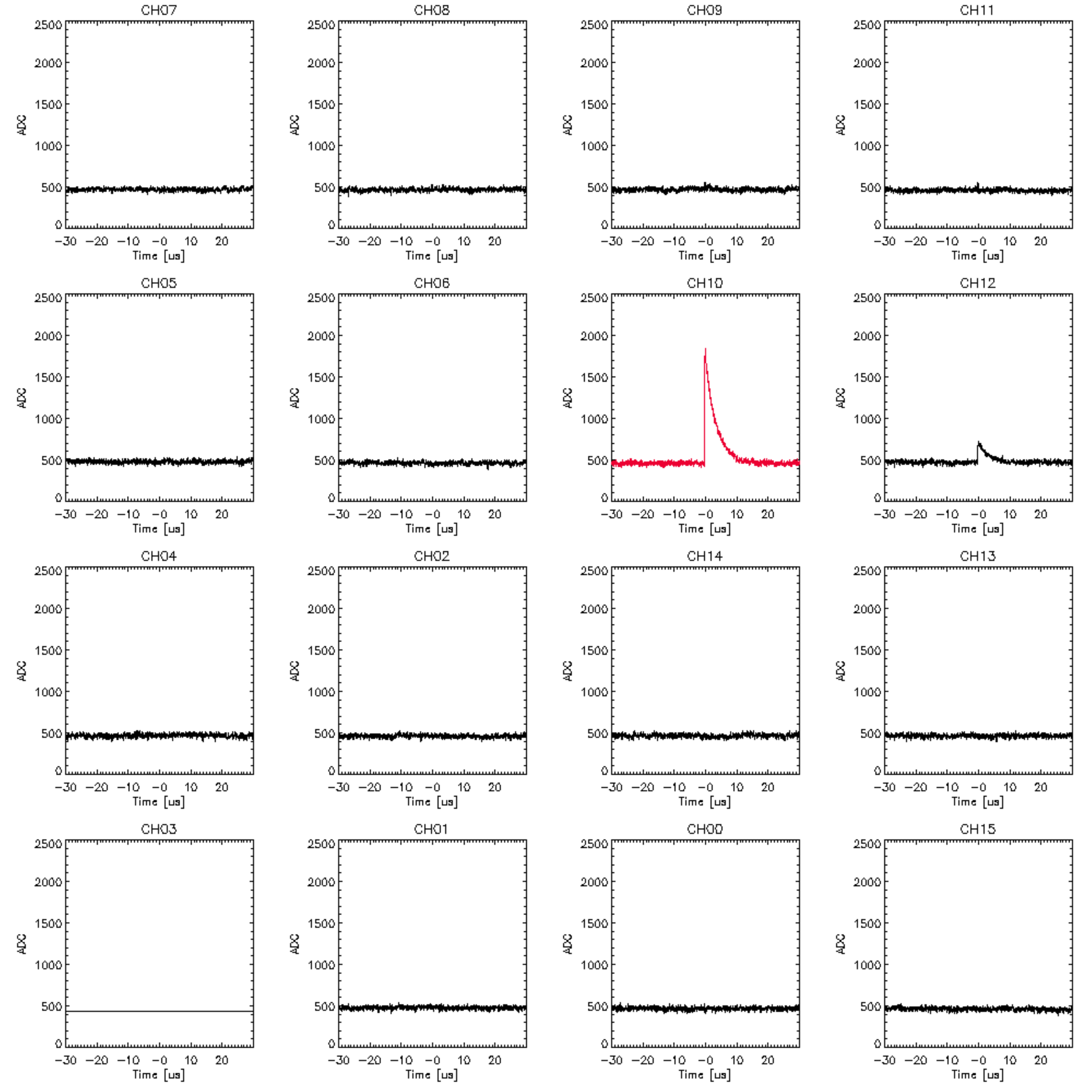}
\caption{Two Mn $\mathrm{K_{\alpha}}$ signals (5.9~keV) acquired with the multi-channel fast digitization system. Top panels show a single event ($m = 1$) waveform while bottom panels show a charge-shared event ($m = 2$) between CH10 and CH12.}
\label{fig:sharing}       
\end{figure}

\begin{figure}[!t]
\centering
\includegraphics[width=0.9\textwidth]{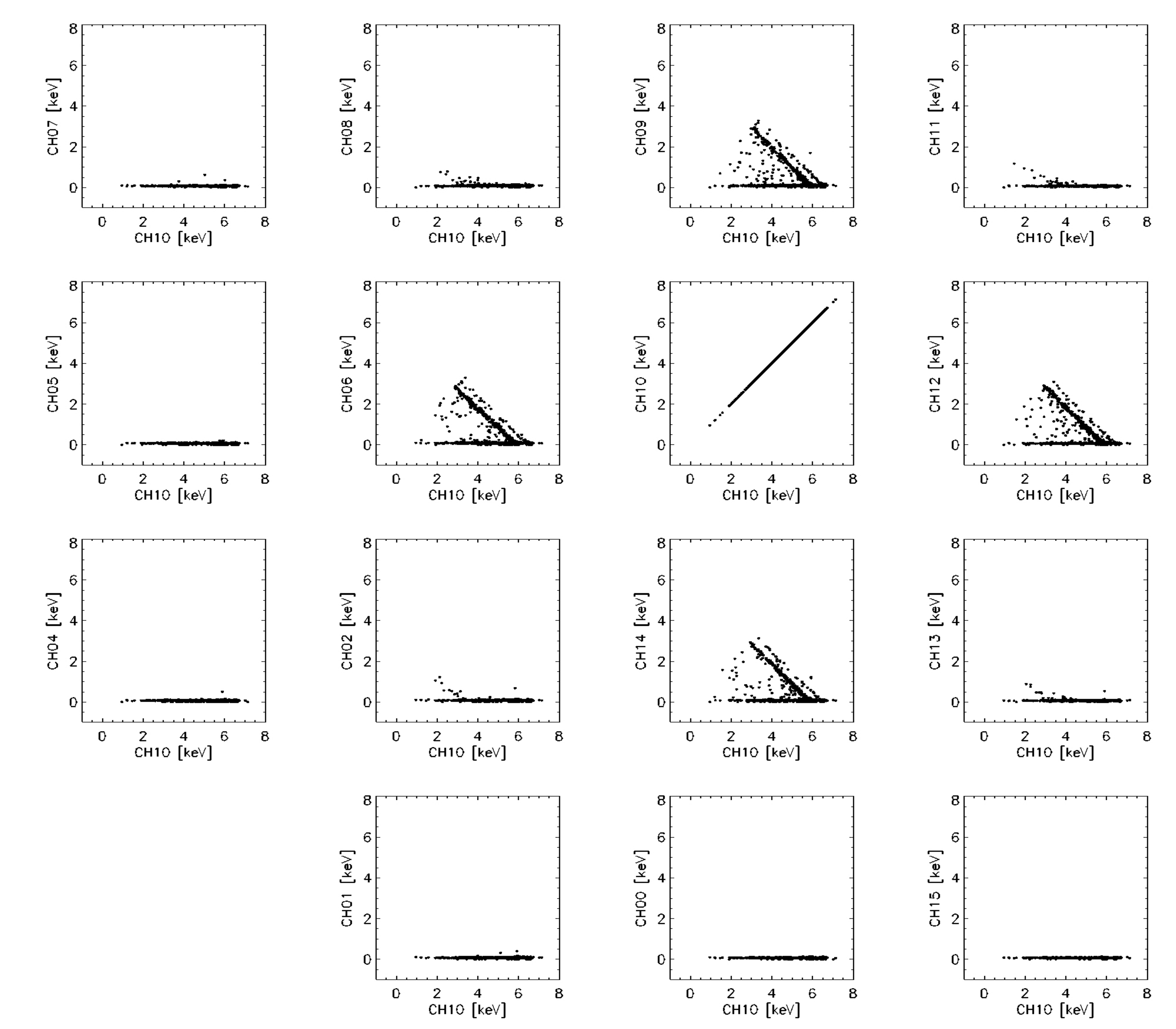}
\caption{Distribution of the charge collected  by CH10 (X-axis) versus the charge measured by the other pixels (Y-axis). Measurement points distributed parallel to the abscissa axis show single events (i.e. no charge sharing), while the points distributed following approximately the relation $\mathrm{X+Y=K}$ represent shared events.}
\label{fig:scatter}       
\end{figure}

A further analysis was performed to evaluate the presence of charge-shared events between neighbouring pixels. Although due to ionization the initial charge cloud lateral dimension is limited to a few microns, diffusion during charge drift in the silicon bulk is expected to lead to non-negligible cloud broadening. In these conditions, an appreciable fraction of charge-shared events can be observed, that can lead to a sensible worsening of the detector spectroscopic performance.
For this characterization campaign a non-collimated $\mathrm{^{55}Fe}$ X-ray source was used to simultaneously illuminate the whole PixDD detector.
In Figure~\ref{fig:sharing} (top and bottom panels) two signals of 5.9~keV ($\mathrm{Mn\,K_{\alpha}}$) are shown. Top panels show the waveforms acquired by all the channels when a single event (multiplicity $m = 1$) is detected. For this event, the charge produced during the photon interaction was entirely collected by one channel (CH10). For comparison, bottom panels show a charge-shared event ($m = 2$), for which the charge cloud has been collected by two adjacent channel (CH10 and CH12).
In order to quantitatively evaluate the amount of charge sharing between neighbouring channels, we analysed the dependence of the charge collected by the 4 central pixels with the signal of the 8 relative neighbouring pixels. In the following we present the analysis performed on CH10, but consistent results were obtained for the four central pixels.
The acquired events were selected following the condition that more than 50\% of the total charge was collected by CH10, this to ensure that the photon interaction took place in CH10 volume.

In Figure~\ref{fig:scatter} each panel shows the distribution of the charge collected  by CH10 (X-axis) versus the charge measured by the other pixels (Y-axis). 
Measurement points distributed parallel to the abscissa axis show single events (i.e. no charge sharing), while the points distributed following approximately the relation $\mathrm{X+Y=K}$ represent shared (double) events. From the plots is clear that a fraction of the events experienced charge sharing, and that sharing between more than two nearby pixels is less probable. Still a few triple-events are present, as it can been seen from the presence of measurement points not following the $\mathrm{Y=K}$ and the $\mathrm{X+Y=K}$ relations and from the signal distribution on CH08, CH11, CH02 and CH13. 

\begin{figure}[!t]
\centering
\includegraphics[width=0.85\textwidth]{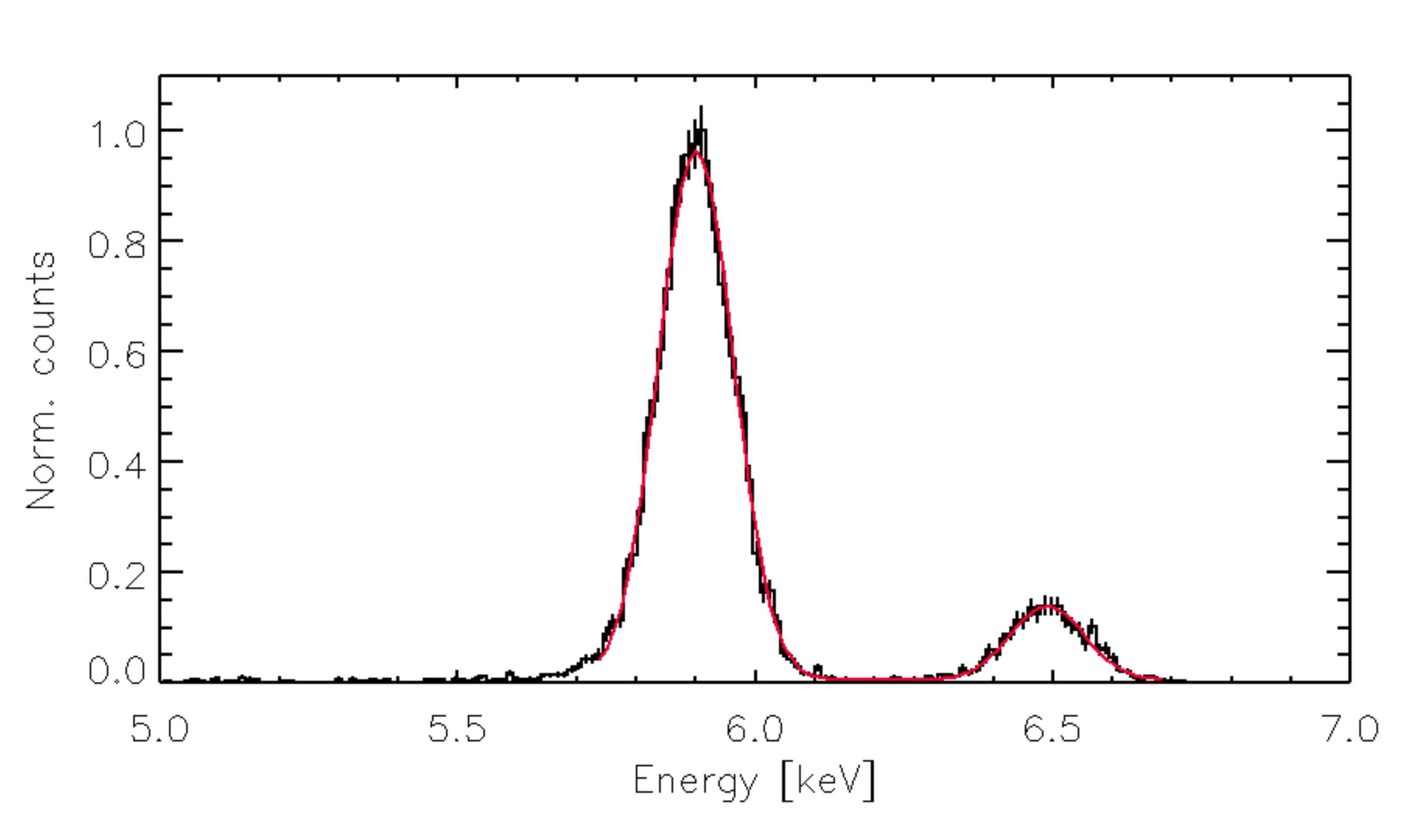}
\caption{
$^{55}$Fe energy spectrum at $\mathrm{+27\,^{\circ}C}$ obtained by summing together single events ($\mathrm{m = 1}$) acquired by CH10 and multiple events ($\mathrm{m \geq 2}$) collected by CH10 and the neighbouring pixels.
}
\label{fig:55ferecon}       
\end{figure}

By setting a threshold equal to 5 times the channel r.m.s noise for the event multiplicity determination, single events ($m = 1$) resulted to be 83.8\% of the total, with an energy resolution of 139.2~eV FWHM, while double and triple ($m = 2$ and $m = 3$) events were quantified to be 15.8\% and  0.4\% respectively.
When properly reconstructed, single and multiple events lead to an overall CH10 energy resolution of 149.8~eV FWHM at 5.9~keV, as shown in Figure~\ref{fig:55ferecon}.



\section{Conclusion and future perspectives}
\label{sec:conclusion}

In the framework of the RedSoX Italian collaboration, a novel, pixelated, silicon drift detector (PixDD) has been designed and produced to fill the technological gap in the capability of realizing pixel silicon detectors with fast response, which are capable of asynchronously detect and trigger on the individual X-ray photons and can offer a nearly Fano-limited energy resolution and soft X-ray response at room temperature, as required by high throughput X-ray astronomy applications.

The whole system, composed of a 16 pixels $\mathrm{500\,\mu m\,\times \,500 \,\mu m}$ wide, has been integrated with ultra low-noise front-end electronics (SIRIO 3.4) and characterized at $\mathrm{+ 27\,^{\circ}C}$ at the INAF-IAPS laboratories, demonstrating outstanding noise figure (single channel $\mathrm{ENC \leq 10\, e^{-}}$ r.m.s.) and spectroscopic capabilities ($\mathrm{\Delta E \leq 150\,eV}$ FWHM at 5.9~keV).
The physical and electronic behaviour of the individual pixels have been studied to assess the charge sharing between neighbouring detector cells, resulting in a limited fraction of shared events ($\simeq$16\%) whose contribution to the energy resolution does not undermine the overall detector performance.

Further experimental activities have already taken place aimed at the design and production of a large PixDD detector, composed by $\mathrm{16 \times 8}$ pixels with $\mathrm{300 \times 300\,\mu m^2}$ dimensions. 
The $\mathrm{16 \times 8}$ PixDD will be read-out by a monolithic $16\times 8$ channel ASIC (namely RIGEL) developed by Polytechnic of Milan and University of Pavia and currently in production.  The PixDD--RIGEL hybrid will be flip-chip-bonded exploiting the bump bonding technique developed at the Karlsruhe Institute of Technology (KIT) \citep{Caselle2016}.
\\

\acknowledgments

The authors warmly acknowledge support by INFN (under the RedSoX2 project and FBK-INFN agreement 2015-03-06), INAF (under grant TECNO-INAF-2014) and ASI (under  INAF-ASI agreement 2016-18-H.0).

\bibliographystyle{JHEP}
\bibliography{biblio}

\providecommand{\href}[2]{#2}\begingroup\raggedright\begin{thebibliography}{10}

\bibitem{Kemmer1987}
J.~{Kemmer} and G.~{Lutz}, \emph{{New detector concepts}},
  \href{http://dx.doi.org/10.1016/0168-9002(87)90518-3}{\emph{Nuclear
  Instruments and Methods in Physics Research A} {\bfseries 253} (Jan., 1987)
  365--377}.

\bibitem{Kemmer1990}
J.~{Kemmer}, G.~{Lutz}, U.~{Prechtel}, K.~{Schuster}, M.~{Sterzik},
  L.~{Str{\"u}der} et~al., \emph{{Experimental confirmation of a new
  semiconductor detector principle}},
  \href{http://dx.doi.org/10.1016/0168-9002(90)90470-Q}{\emph{Nuclear
  Instruments and Methods in Physics Research A} {\bfseries 288} (Mar., 1990)
  92--98}.

\bibitem{Lutz2001}
G.~{Lutz}, R.~H. {Richter} and L.~{Str{\"u}der}, \emph{{Novel pixel detectors
  for X-ray astronomy and other applications}},
  \href{http://dx.doi.org/10.1016/S0168-9002(00)01258-4}{\emph{Nuclear
  Instruments and Methods in Physics Research A} {\bfseries 461} (Apr., 2001)
  393--404}.

\bibitem{Meidinger2017}
N.~{Meidinger}, M.~{Barbera}, V.~{Emberger}, M.~{F{\"u}rmetz}, M.~{Manhart},
  J.~{M{\"u}ller-Seidlitz} et~al., \emph{{The Wide Field Imager instrument for
  Athena}},  in \emph{Society of Photo-Optical Instrumentation Engineers (SPIE)
  Conference Series}, vol.~10397 of \emph{SPIE Conference Series}, p.~103970V,
  Aug., 2017.
\newblock \href{http://dx.doi.org/10.1117/12.2271844}{DOI}.

\bibitem{Porro2012}
M.~{Porro}, L.~{Andricek}, S.~{Aschauer}, M.~{Bayer}, J.~{Becker},
  L.~{Bombelli} et~al., \emph{{Development of the DEPFET Sensor With Signal
  Compression: A Large Format X-Ray Imager With Mega-Frame Readout Capability
  for the European XFEL}},
  \href{http://dx.doi.org/10.1109/TNS.2012.2217755}{\emph{IEEE Transactions on
  Nuclear Science} {\bfseries 59} (Dec., 2012) 3339--3351}.

\bibitem{Bertuccio2007}
G.~{Bertuccio} and S.~{Caccia}, \emph{{Progress in ultra-low-noise ASICs for
  radiation detectors}},
  \href{http://dx.doi.org/10.1016/j.nima.2007.04.042}{\emph{Nuclear Instruments
  and Methods in Physics Research A} {\bfseries 579} (Aug., 2007) 243--246}.

\bibitem{Bertuccio2014}
G.~Bertuccio, D.~Macera, C.~Graziani and M.~Ahangarianabhari, \emph{A cmos
  charge sensitive amplifier with sub-electron equivalent noise charge},  in
  \emph{2014 IEEE Nuclear Science Symposium and Medical Imaging Conference
  (NSS/MIC)}, pp.~1--3, Nov, 2014.
\newblock \href{http://dx.doi.org/10.1109/NSSMIC.2014.7431123}{DOI}.

\bibitem{Bertuccio2015}
G.~{Bertuccio}, M.~{Ahangarianabhari}, C.~{Graziani}, D.~{Macera}, Y.~{Shi},
  A.~{Rachevski} et~al., \emph{{A Silicon Drift Detector-CMOS front-end system
  for high resolution X-ray spectroscopy up to room temperature}},
  \href{http://dx.doi.org/10.1088/1748-0221/10/01/P01002}{\emph{Journal of
  Instrumentation} {\bfseries 10} (Jan., 2015) P01002}.

\bibitem{Bertuccio2016}
G.~{Bertuccio}, M.~{Ahangarianabhari}, C.~{Graziani}, D.~{Macera}, Y.~{Shi},
  M.~{Gandola} et~al., \emph{{X-Ray Silicon Drift Detector-CMOS Front-End
  System with High Energy Resolution at Room Temperature}},
  \href{http://dx.doi.org/10.1109/TNS.2015.2513602}{\emph{IEEE Transactions on
  Nuclear Science} {\bfseries 63} (Feb., 2016) 400--406}.

\bibitem{Bufon2018}
J.~{Bufon}, S.~{Schillani}, M.~{Altissimo}, P.~{Bellutti}, G.~{Bertuccio},
  F.~{Bill{\`e}} et~al., \emph{{A new large solid angle multi-element silicon
  drift detector system for low energy X-ray fluorescence spectroscopy}},
  \href{http://dx.doi.org/10.1088/1748-0221/13/03/C03032}{\emph{Journal of
  Instrumentation} {\bfseries 13} (Mar., 2018) C03032}.

\bibitem{Spieler2005}
H.~Spieler, \emph{Semiconductor detector systems}.
\newblock Oxford Science Publications, Oxford U.K., 2005.

\bibitem{Guzik2013}
Z.~{Guzik} and T.~{Krakowski}, \emph{{Algorithms for digital gamma ray
  spectroscopy}}, {\emph{NUKLEONIKA} {\bfseries 58} (2013) 333--338}.

\bibitem{Muleri2008}
F.~{Muleri}, P.~{Soffitta}, R.~{Bellazzini}, A.~{Brez}, E.~{Costa}, M.~{Frutti}
  et~al., \emph{{A versatile facility for the calibration of x-ray polarimeters
  with polarized and unpolarized controlled beams}},  in \emph{Space Telescopes
  and Instrumentation 2008: Ultraviolet to Gamma Ray}, vol.~7011 of \emph{SPIE
  Conference Series}, p.~701127, July, 2008.
\newblock \href{http://dx.doi.org/10.1117/12.789605}{DOI}.

\bibitem{Caselle2016}
M.~{Caselle}, T.~{Blank}, F.~{Colombo}, A.~{Dierlamm}, U.~{Husemann},
  S.~{Kudella} et~al., \emph{{Low-cost bump-bonding processes for high energy
  physics pixel detectors}},
  \href{http://dx.doi.org/10.1088/1748-0221/11/01/C01050}{\emph{Journal of
  Instrumentation} {\bfseries 11} (Jan., 2016) C01050}.

\end{thebibliography}\endgroup

\end{document}